# A note on the two-neutrino decay mode of excited nuclear states

I.V.Aničin

*Faculty of Physics, University of Belgrade, Belgrade, Serbia*

**Abstract**: The possibility for the excited nuclear states to decay by emission of the neutrino-antineutrino pair in direct competition with electromagnetic decay modes is commented upon.

Elementary considerations suggest that the existence of weak neutral current should provide for yet another decay channel of excited nuclear states. That it has up to now neither been suspected nor observed, and that it will hardly be observed in the future, is firstly due to its low probability and secondly due to its practically unobservable decay products. The process would run via the emission of the virtual neutral intermediate $Z^0$ boson that would in turn decay into the mixture of real neutrino-antineutrino pairs, of all flavors allowed by mass-energy conservation. We shall call this the 2ν decay. Its probability is intrinsically of the order of the probability for beta decays, or for the emission of the virtual charged W boson that subsequently decays into the real electron-neutrino lepton pair. However, the 2ν decay mode always competes with the many orders of magnitude more probable electromagnetic decay modes, and is, minding the difficulties with the detection of low energy neutrinos, practically unobservable even in highly forbidden isomeric transitions, where the electromagnetic modes are suppressed. As an example we consider the decay of the 388 keV isomeric state in $^{87}$Sr, which is a well known case of the cascade of beta decays that compete successfully with the electromagnetic decay (Fig.1) (see [1]). This succession of weak decays shares the same initial and final states with the electromagnetic and 2ν modes and effectively results in the emission of the electron neutrino-antineutrino pair, though it is realized via the two charged current interactions through a real intermediate state, and not by a single neutral current interaction, denoted by 2ν and marked by the dashed arrow in the drawing. The 2ν mode should here have a practically unobservable branching ratio smaller than 2.8h/4.75×10$^{10}$y, due to the spin difference which is necessarily higher than that for the already highly forbidden beta decay of $^{87}$Rb, for which the log $ft$ value equals 17.6. Since it is not very likely that the more favorable case is supplied by nature, the 2ν decay mode is probably bound to remain just another curiosity in the nuclear zoo.

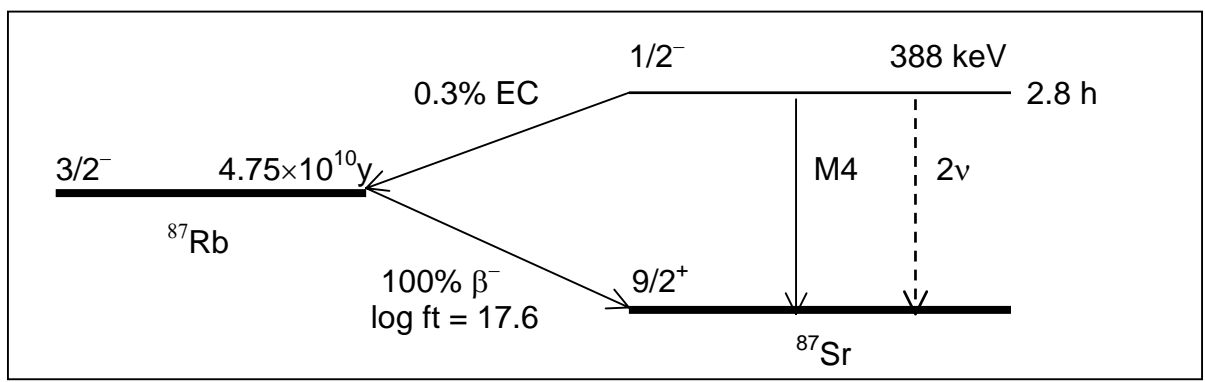

**Fig.1.** The decay scheme of $^{87m}$Sr, with the 2ν decay mode marked by the dashed arrow, which is equivalent to the EC-β$^-$ succession of decays, as discussed in the text.